\documentclass[12pt]{article}
\usepackage{graphicx,amsmath,amssymb,cite}
\usepackage{epsfig,epsf}
\usepackage{graphicx}
\usepackage{epstopdf}
\newcommand{\beq}{\begin{equation}}
\newcommand{\eeq}{\end{equation}}

\numberwithin{equation}{section}

\begin{document}

\begin{flushright}
\end{flushright}
\vspace*{0.5 cm}

\begin{center}

{\Large{\bf Investigation of  High Energy Behaviour  \\  of  
HERA Data  }}

\vspace*{1 cm}

{\large {A. Luszczak~$^1$ and H. Kowalski~$^2$ }} \\ [0.5cm]
{\it $^1$ T.Kosciuszko, Cracow University of Technology, Institute of Physics, st. Podchorazych 1, 30-084 Krakow, Poland}\\[0.1cm]
{\it $^2$ Deutsches Elektronen-Synchrotron DESY, Berlin and Hamburg, Germany}\\[0.1cm]
 \end{center}

\vspace*{3 cm}

\begin{center}
{\bf Abstract}  \end{center}
We analyse the high precision HERA $F_2$ data in the low-$x$, $x<0,01$, and  very-low-$x$, $x<0.001$, regions  using $\lambda$-fits.   $\lambda$  is a measure of the rate of rise of $F_2$ defined by $F_2 \propto (1/x)^{\lambda}$. We show that $\lambda$ determined in these two regions, at various  $Q^2$ values, is systematically smaller in the very-low-$x$ region as compared to the low-$x$ region. We discuss some possible  physical interpretations of this effect.


\vspace*{3 cm}


\newpage

\section{Introduction}

 From the first measurements of HERA it was already known,   that the rise of $F_2$ with diminishing $x$ can be described by a simple parametrisation; 
$F_2 \propto(1/x)^\lambda$, where the parameter $\lambda$ is a function of $Q^2$ \cite{BartKow2001}.
The observed values of  $\lambda$ are increasing with $Q^2$, at $Q^2= 0.35 $ GeV$^2$   $\lambda$ is about 0.1 and at $Q^2= 250 $ GeV$^2$ 
 $\lambda $ is about 0.4.  

Before HERA, it was expected that the virtual photon-hadron cross section, $\sigma^{\gamma^* p}$, should rise with energy in a similar way as the hadronic cross sections. The value of the exponent $\lambda$, obtained from the analysis of hadron-hadron scattering data, was about 0.08 and it was expected that this value should be universal,  i.e. independent of the type of reaction and of $Q^2$ \cite{DonLan}.  The physical motivation behind  this expectation relied on the fact that in the low-$x$ region the interaction time is very long (in the proton rest frame it is given by $1/(m_p x$)) and therefore the incoming virtual photon has enough time to get "dressed", i.e. to behave like a hadron.  

The observation of    $\lambda$ values, which were sizeably larger than 0.08 and were rising with $Q^2$  was a clear sign that $F_2$ is measuring a fundamental partonic process, since in QCD the phase space for gluon emission grows  quickly  with increasing $Q^2$ and decreasing $x$.  One of the possible physical interpretations of the parameter $\lambda$ is that it is a rate of gluon emissions per unit of rapidity, see e.g.  \cite{BartKow2001}.

Long before HERA,  it was known from QCD, that the gluon density, in the double asymptotic limit of $x \rightarrow 0$ and $Q^2 \rightarrow \infty $,   should behave as 
$$ xg(x,Q^2) = \exp \sqrt{ C  \ln \frac{\ln Q^2/\Lambda}{\ln Q^2_0/\Lambda}\,\ln1/x} $$   
where $C$ is a know constant and $\Lambda$ is the QCD scale \cite{DeRujula,DGLAP}, i.e.  $F_2$  grows with decreasing $x$ and increasing $Q^2$. 
The same double asymptotic limit is also derived in BFKL \cite{BFKL}, see e.g. the discussion in \cite{KLR16}. 

 In the HERA region, i.e. well below the double asymptotic domain, the data can be well described,
including the $\lambda$ behaviour,    by fitting  the starting  density distributions of quarks and gluons with DGLAP. The final, high precision HERA data has shown, however, that something is missing in these fits \cite{hiH1ZEUS}. There is now a growing consensus that the observed discrepancies are due to the BFKL effects in the low-$x$ region, where DGLAP evolution should not be valid, see e.g. \cite{Ball}.

Before HERA,  the standard prediction of BFKL  was that $ F_2$,  in the low-$x$ region, should be dominated by the leading singularity, i.e $F_2 \propto (1/x)^\omega$, with $\omega \approx 0.3$ (in NLO) \cite{RosFor,salam}. This prediction, obtained from the evaluation of the BFKL amplitude in the Mellin space using the saddle point approximation, was in clear contradiction with the observed increase of the exponent $\lambda$ with $Q^2$. 
In an alternative approach, Lev Lipatov proposed \cite{lipatov86}  to introduce the infrared cutoff, which leads to a discrete BFKL pole structure when the running of the strong coupling constant, $\alpha_s$, is taken into account. It was then shown that the combined effect of these poles reproduces the observed behaviour of the parameter $\lambda$
  \cite{KLRW}    and provides an excellent fit to the high precision  HERA data, in the very low $x$ region, $x<0.001$ and $Q^2> 6 $ GeV$^2$ \cite{KLRS17}.  

The investigation of ref.  \cite{KLRS17} led to an unexpected result:  the $F_2$ data were described by the sub-leading poles only. Moreover, it was found that the leading pole  has to be in the saturation or  even  non-perturbative region, so that its contribution to the investigated region had to be negligible. The  second pole, which is leading for $Q^2>6$ GeV$^2$, has  $\omega \approx 0.14$, i.e. a substantially lower value than the values of the observed  $\lambda$'s in this $Q^2$ region. This result leads to the expectation that at very low $x$ one could observe a substantially lower values of  $\lambda$. 

Another interesting feature of HERA data was pointed out in ref.  \cite{CaldWing},  where it is shown that  the logarithmic linearity of HERA data,  $\log F_2 \propto \lambda \log (1/x)$,  has to break down.  The argument is simple; the  ${\gamma p}$ cross section, $\sigma^{\gamma p} = 4\pi \alpha F_2 / Q^2 $, is smaller for larger $Q^2$ but 
it  rises faster with diminishing $x$. Therefore, there is a cross over point, at about $x = 10^{-8}$, where the $\gamma p$ cross sections for larger $Q^2$'s are becoming  larger than the cross sections for smaller $Q^2$. This is a contradiction because the cross section of a smaller projectile cannot be larger than the cross section of a larger one (the transverse size of the incoming photon is given by $1/Q^2$).    

All this provides a motivation to refit the HERA high precision data  \cite{hiH1ZEUS} with the  parametrisation, $F_2 \propto(1/x)^\lambda$, using the full set of statistical and correlated systematic errors. We want to see whether the logarithmic linearity  still holds or whether we can see any change in the rate of rise, $\lambda$, with  diminishing $x$.

\section{Description of fits}  					

We determine the  $\lambda$ parameter from a fit of the function $F_2 = A \, (1/x)^\lambda$  to the final, high precision H1 and ZEUS data, measured at HERA \cite{hiH1ZEUS}. Here, $A$ is a normalisation constant dependent on $Q^2$.  The fits are made in the range  $x < 0.01$ and $x<0.001$, for each $Q^2$ value separately, with $0.3 <Q^2  < 250$ GeV$^2$.  The final data  are given as  reduced cross sections, which are connected to $F_2$ and $F_L$ by:
$$ \sigma_{red} \approx F_2 - \frac{y^2}{Y_+} F_L  ,$$
with $Y_+ = 1+(1-y)^2$, where $y$ denotes the inelasticity parameter.
In the selected kinematic region, the contribution of the structure function $F_3$ can be neglected.  Following the discussion of the properties of $F_L$ in the  H1 papers \cite{H1R,H1R2}, we assume  that $F_L$ is proportional to $F_2$. This is also in agreement with the ZEUS results, see e.g.  \cite{Cald,ZEUSFL}.   We have then
$$ F_2 = \frac{\sigma_{red}}{1-\frac{y^2R}{Y_+(1+R) }} ,$$  
where $R= F_L /(F_2-F_L)$. The  evaluation of $F_L$  shows that the ratio $R$ is only weakly dependent on $x$ and $Q^2$, especially in  the kinematic region of the present investigation, $0.01 < y < 0.8$ \cite{H1R2,ZEUSFL}. 
Thus we assumed $R=0.25$, in agreement with data, the QCD DGLAP predictions  and the dipole models  \cite{KowTea,Ewerz,EwNacht,LuszKow}.

The results of the fits are given in Tables 1 and 2.  Table 1 shows the values of  $\lambda$ constant with its error $\delta \lambda$  for two low-$x$ regions, $x<0.01$ and $x<0.001$. Table 2 show the values of  $A$ constant and its errors in the same $x$ regions.  In both Tables 
 we give also the  values of $\chi^2$  of the  individual fits divided by the number of degrees of freedom, $N_{df} $. (The values of $\chi^2$'s and 
 $N_{df} $'s are the same in both Tables as the constants $\lambda$ and $A$ are determined in the same fit.)
  The number of degrees of freedom  is  given by the number of experimental points used in the fit, at a given $Q^2$ value,  lowered by the number of parameters of the fit, i.e. 2.

Table 1 shows that the fits are of good quality, as the average value of  $\chi^2/N_{df} $ is  1.05, at $x<0.01$. The fits in the $x<0.001$ region are of similar quality. In our analysis we concentrate on the region between $2.7 < Q^2 < 45$ GeV$^2$, for which the $F_2$ measurements in the low-$x$ and the very-low-$x$ region overlaps and are different. In this region, the error of $\lambda$, which includes the statistical and correlated systematic uncertainties, is around 2 \% in the low-$x$ and typically twice as big in the very-low-$x$ region.

\begin{figure}[htbp]
\centerline{
\includegraphics[width=15cm,angle=0]{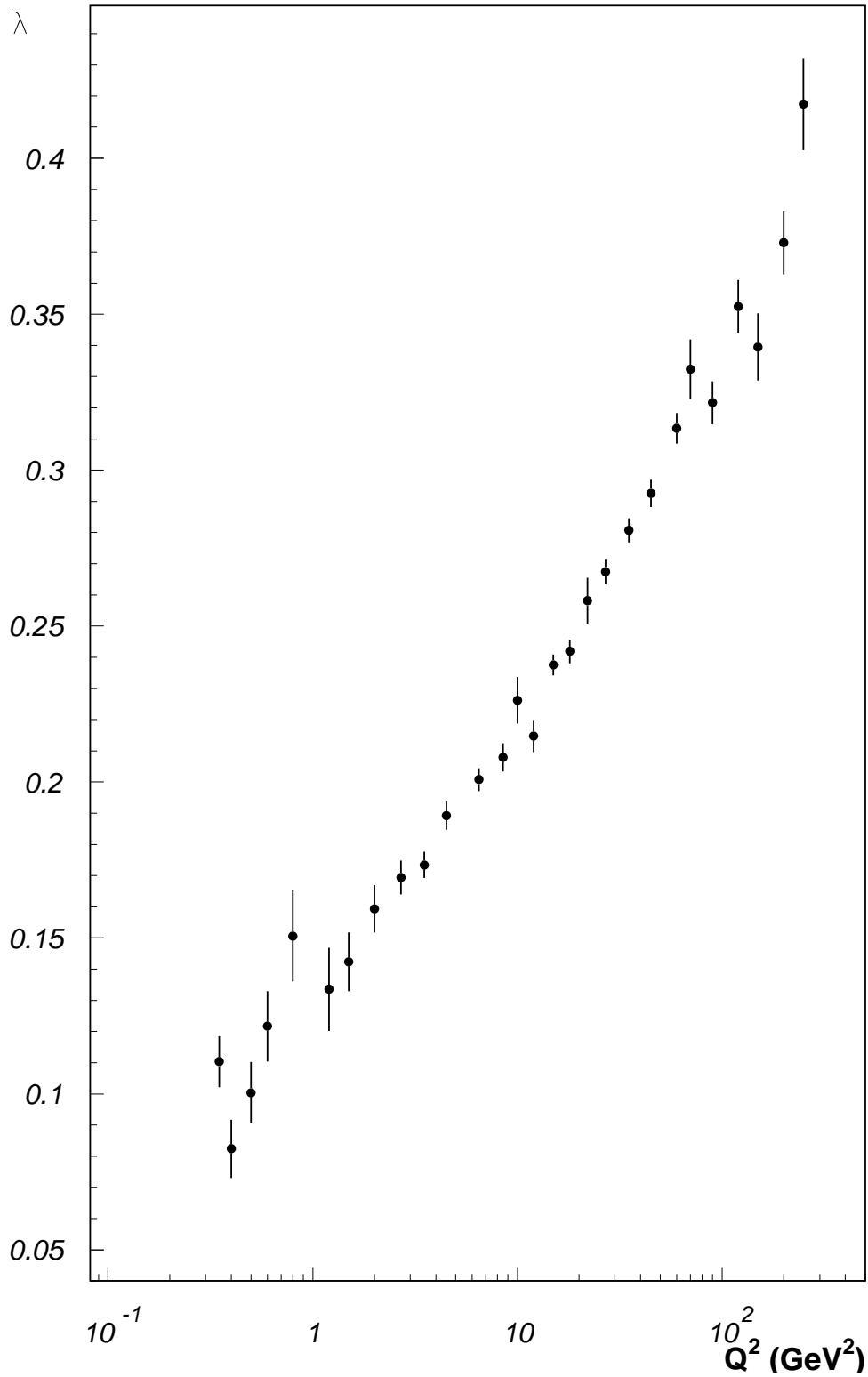}
}
\caption {The power $\lambda$ as determined from the fits  in the  $x<0.01$ region, at all accessible  $Q^2$ values, Table 1.}
\label{lambasic} \end{figure}

\begin{figure}[htbp]
\centerline{
\includegraphics[width=15cm,angle=0]{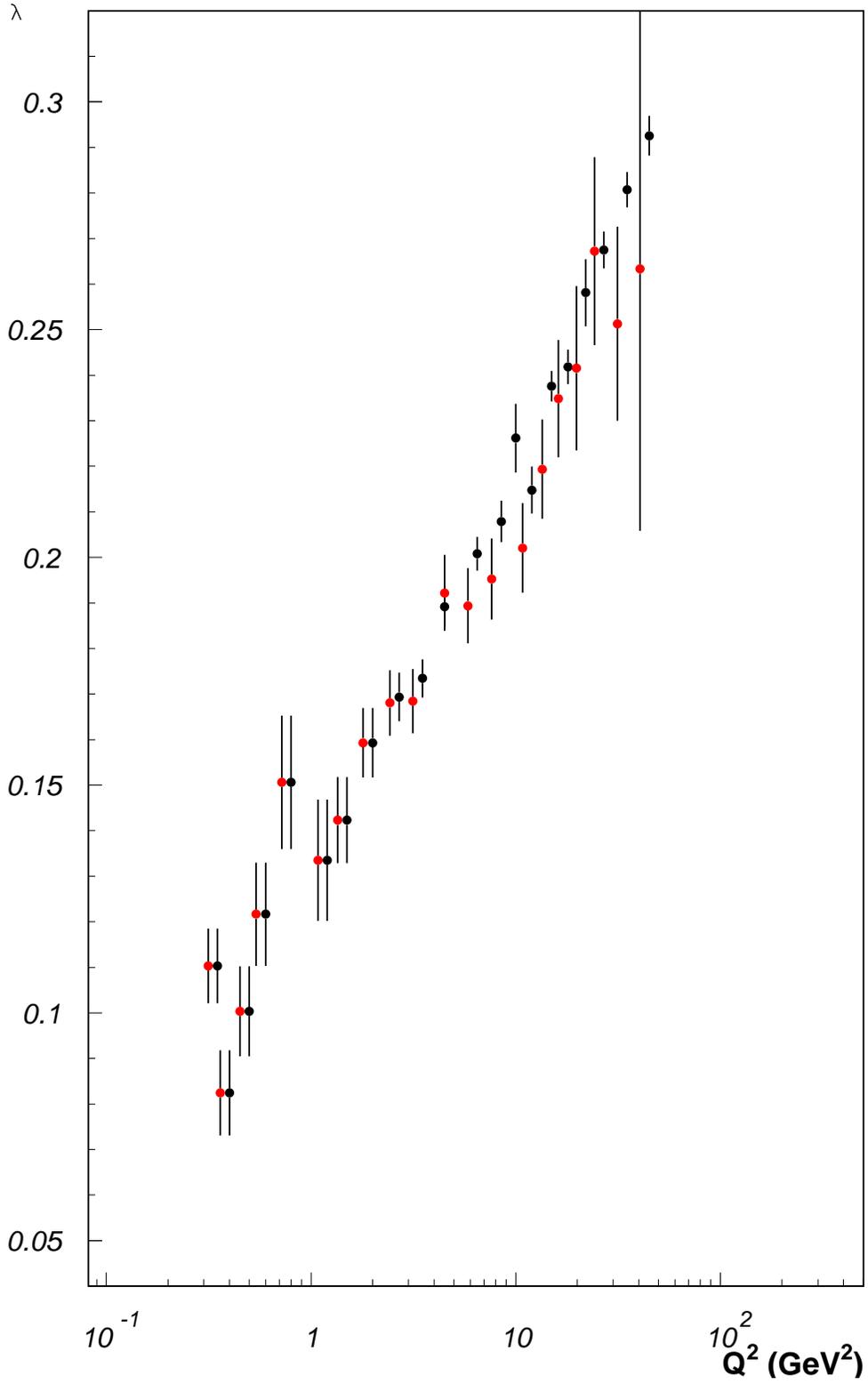}
}
\caption {Comparison of the results of the $\lambda$-fit  in the low-$x$,  $x<0.01$, and very-low-$x$,  $x<0.001$, regions.  Only points determined with differing data are shown, see Table 1. }
\label{lamcomp} \end{figure}

 As is well known, the rate of rise, $\lambda$, increases clearly with the increase of $Q^2$, see Fig.~\ref{lambasic}.  The new observation of this investigation is that the values of $\lambda$ are systematically lower in the  $x<0.001$ region as compared to the  $x<0.01$ region. In Figure~\ref{lamcomp}  we compare the value of $\lambda$ in the two regions and we see that, although the differences for the individual fits are only of the order of one standard deviation,   almost all of the fits show lower values of $\lambda$ in the very low $x$ region.

\begin{figure}[htbp]
\centerline{
\includegraphics[width=9cm,angle=0]{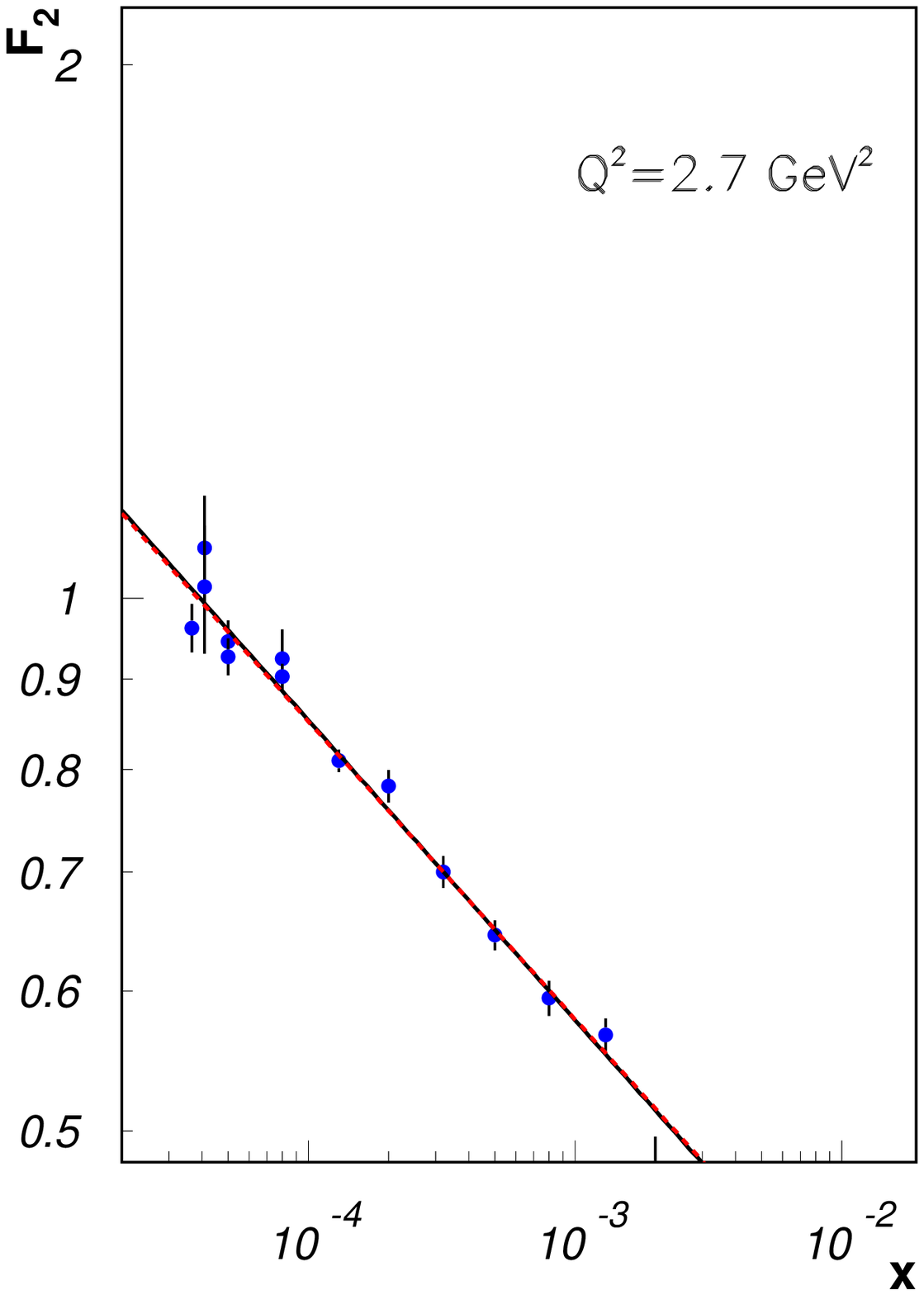}
\includegraphics[width=9cm,angle=0]{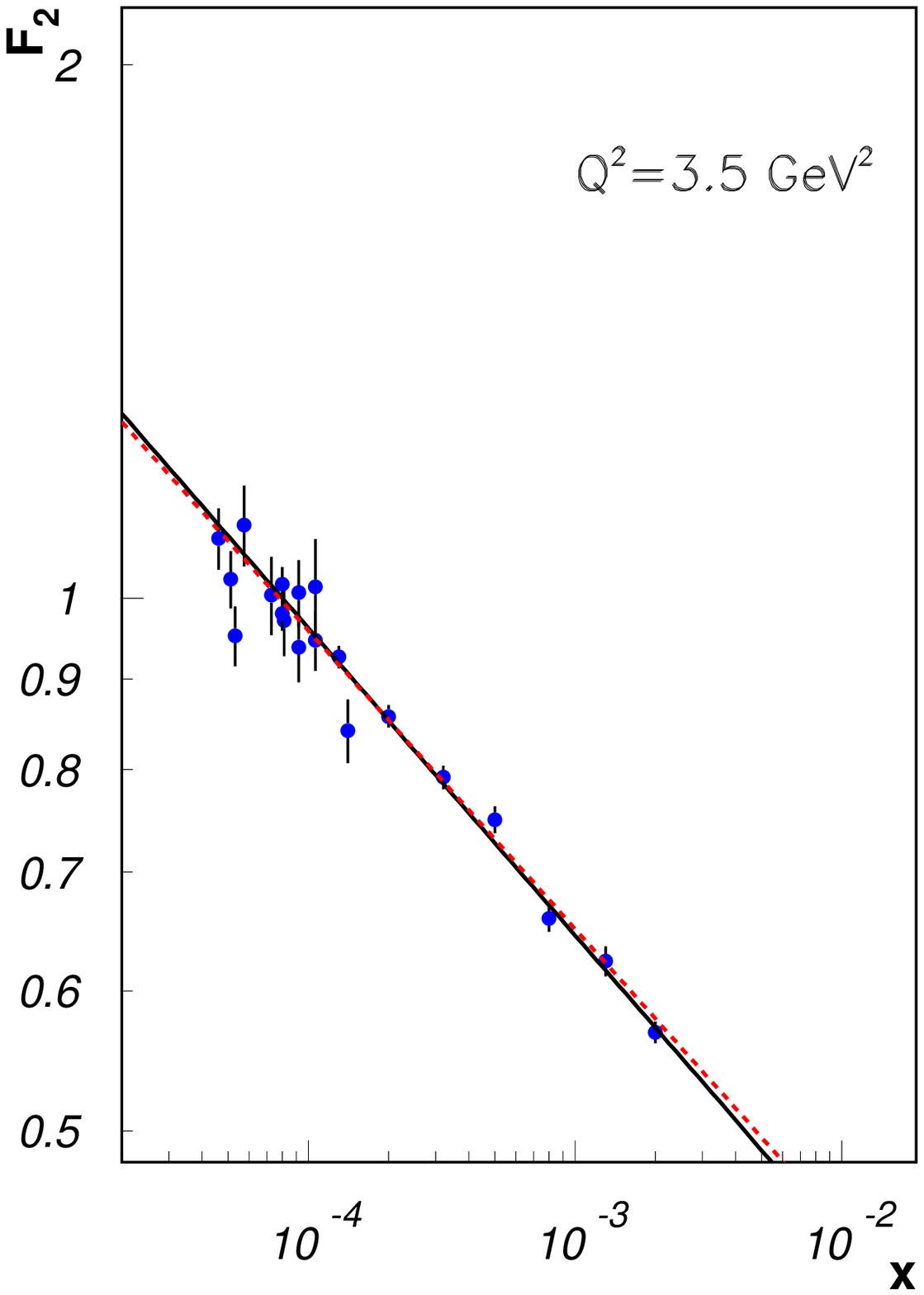}
}
\caption {The fitted curves, $F_2 \propto(1/x)^\lambda$, are shown together with the data at indicated $Q^2$ values. The full line shows the fit in the range $x<0.01$, the dotted line in the range $x<0.001$. The dots show the data with errors bars given by the statistical and systematical errors added in quadrature.  In the fit, all correlations of  errors were taken into account. }
\label{curv1} \end{figure}

\begin{figure}[htbp]
\centerline{
\includegraphics[width=9cm,angle=0]{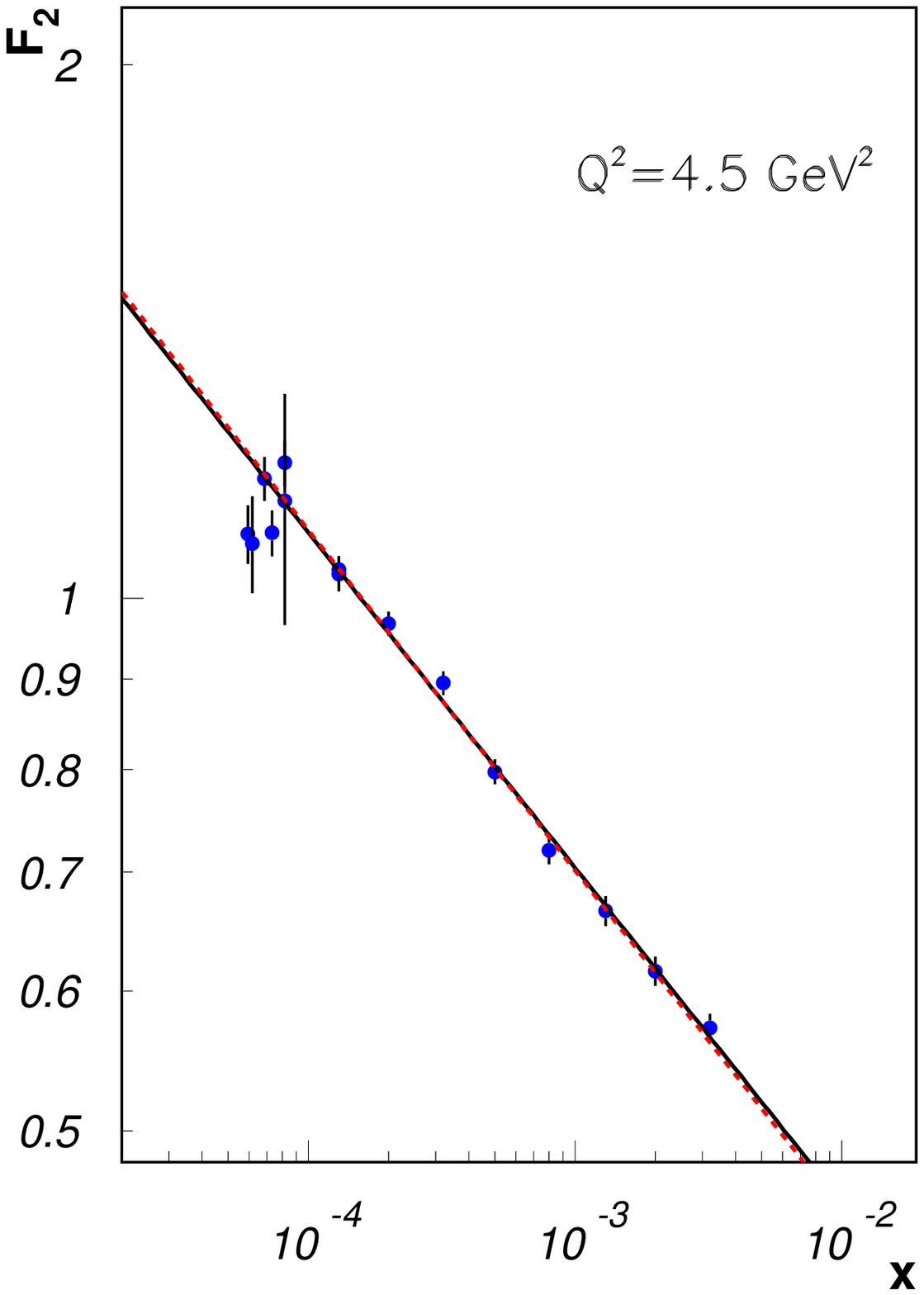}
\includegraphics[width=9cm,angle=0]{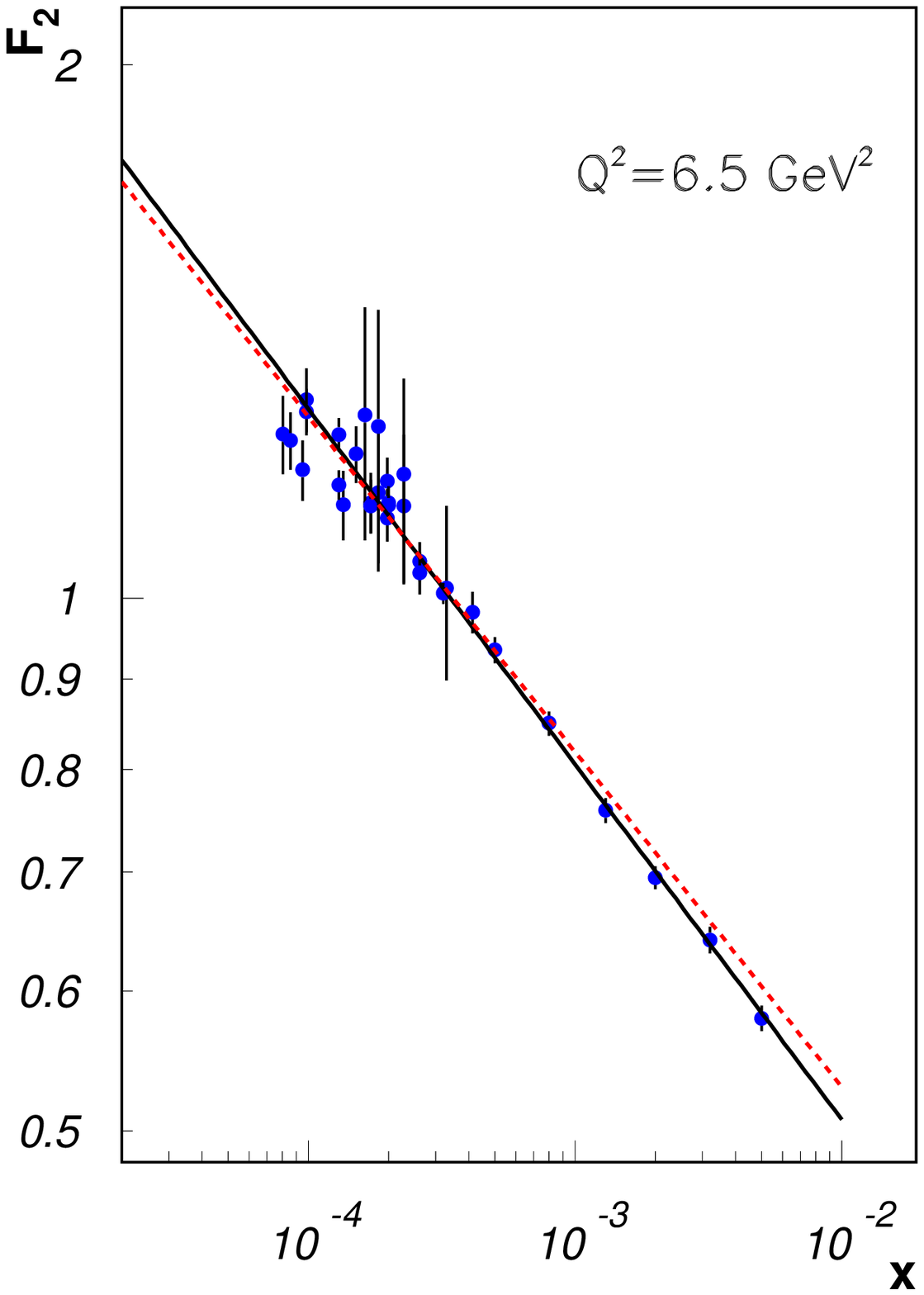}
}
\caption {The fitted curves, $F_2 \propto(1/x)^\lambda$, are shown together with the data at indicated $Q^2$ values. The full line shows the fit in the range $x<0.01$, the dotted line in the range $x<0.001$. The dots show the data with errors bars given by the statistical and systematical errors added in quadrature.  In the fit, all correlations of errors were taken into account. }
\label{curv2} \end{figure}

\begin{figure}[htbp]
\centerline{
\includegraphics[width=9cm,angle=0]{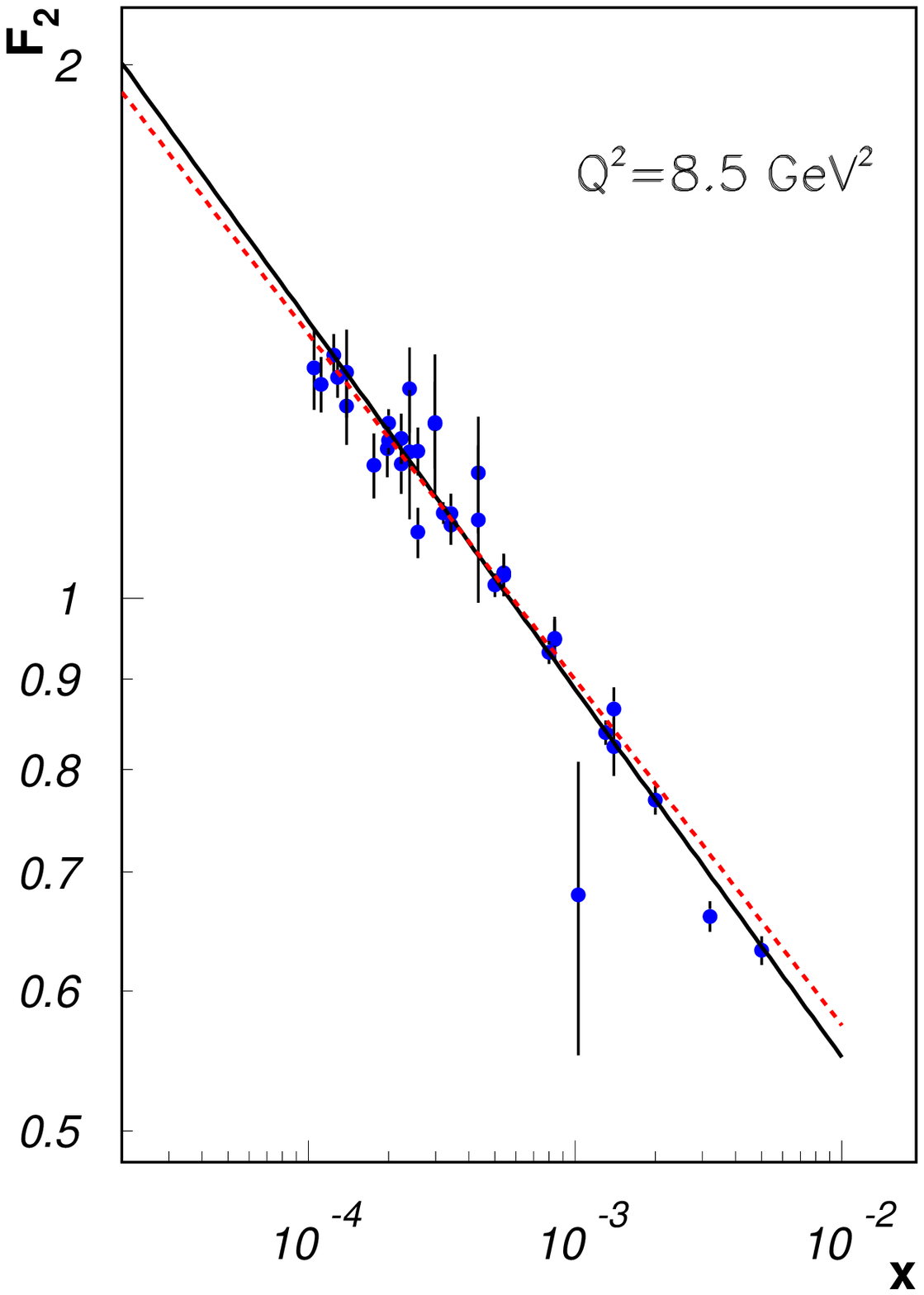}
\includegraphics[width=9cm,angle=0]{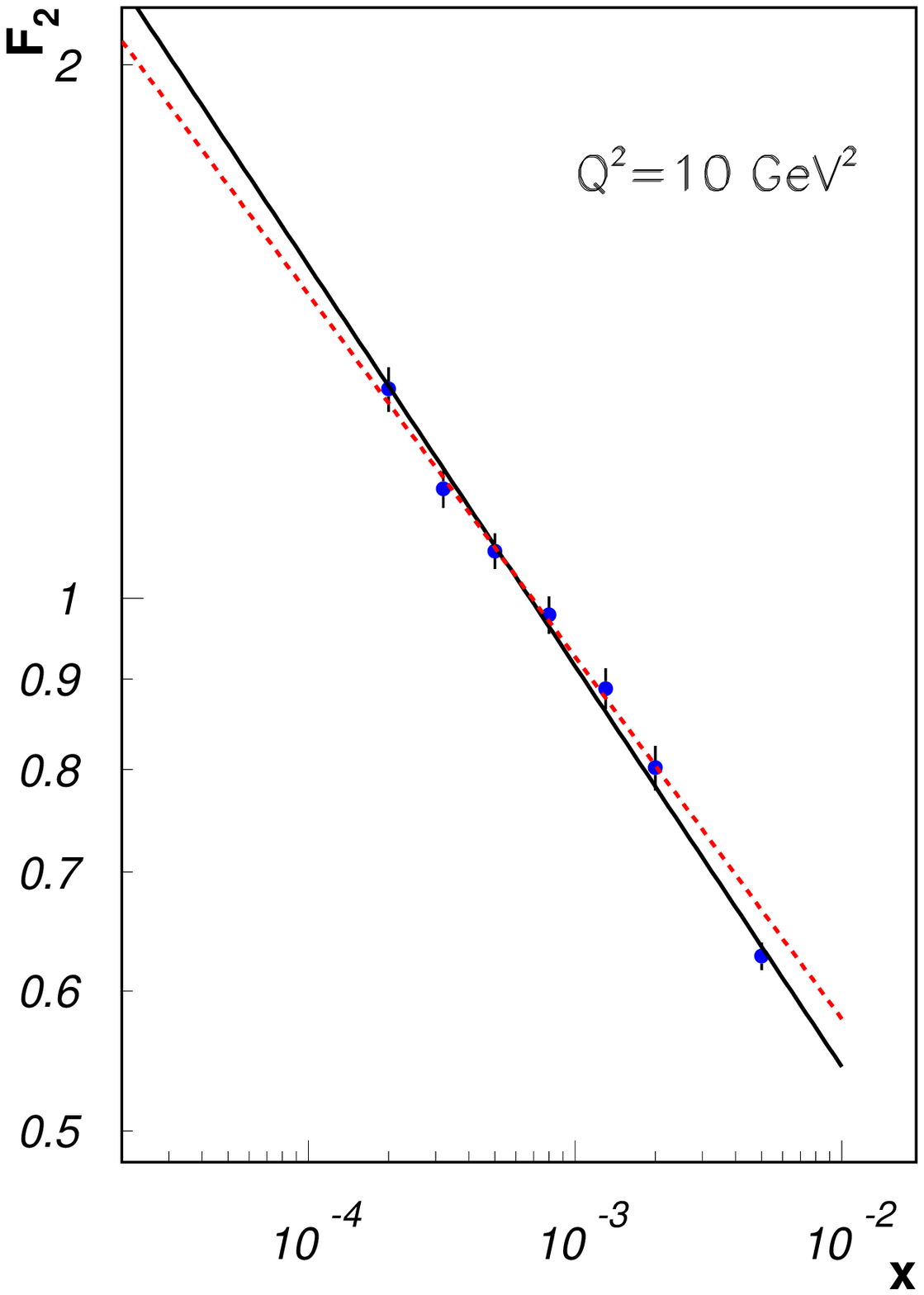}
}
\caption {The fitted curves, $F_2 \propto(1/x)^\lambda$, are shown together with the data at indicated $Q^2$ values. The full line shows the fit in the range $x<0.01$, the dotted line in the range $x<0.001$. The dots show the data with errors bars given by the statistical and systematical errors added in quadrature.  In the fit, all correlations of  errors were taken into account. }
\label{curv3} \end{figure}

\begin{figure}[htbp]
\centerline{
\includegraphics[width=9cm,angle=0]{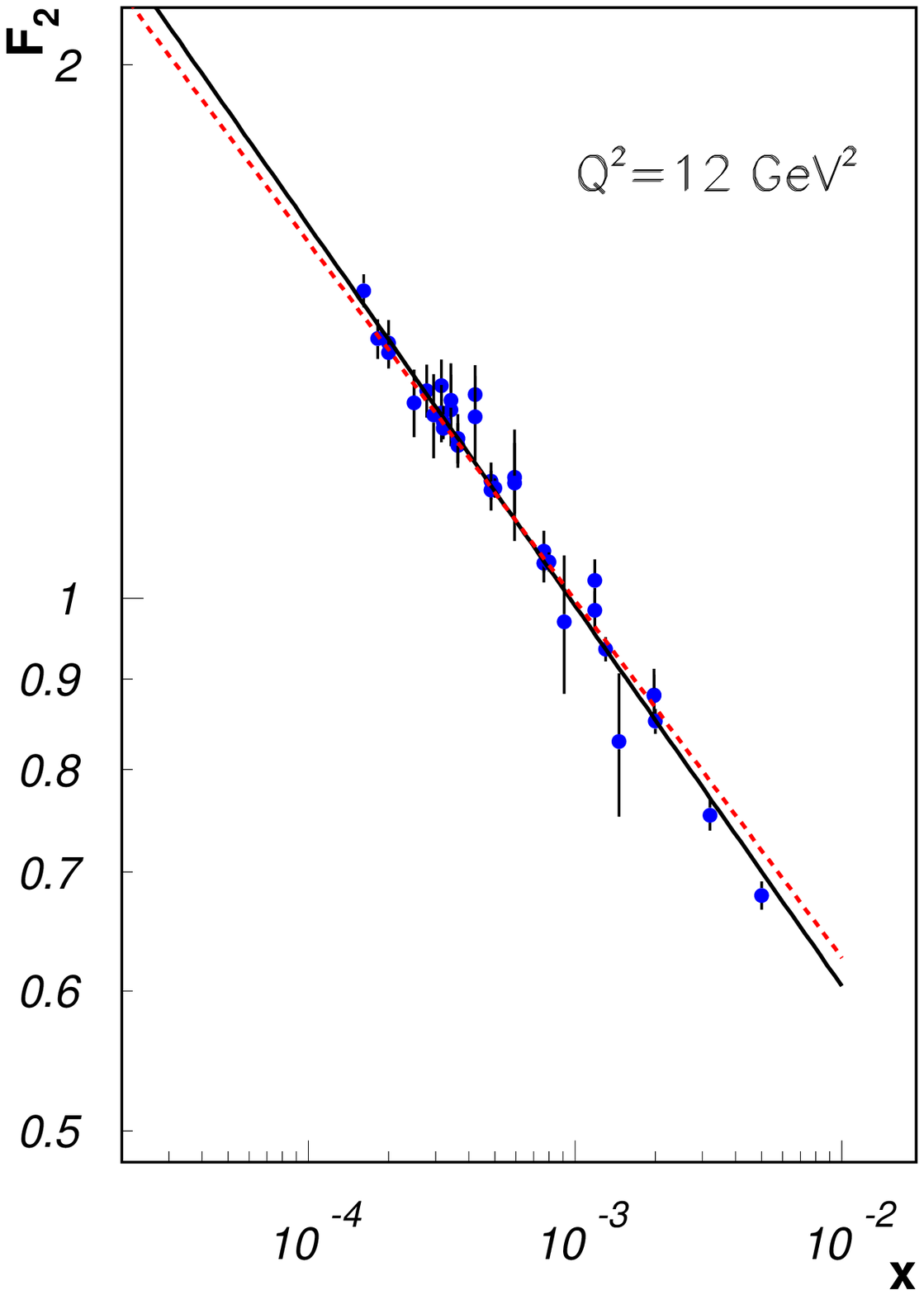}
\includegraphics[width=9cm,angle=0]{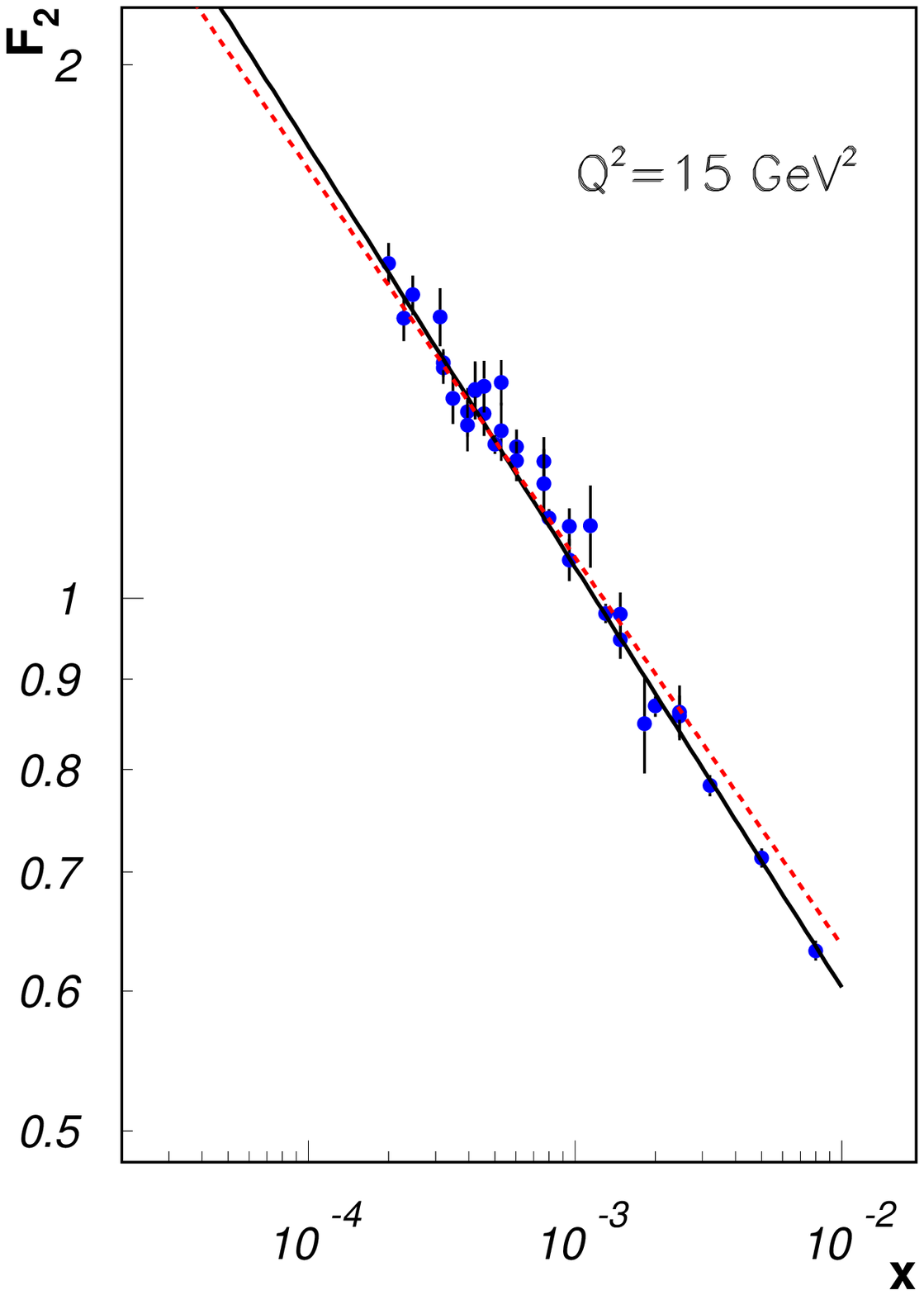}
}
\caption {The fitted curves, $F_2 \propto(1/x)^\lambda$, are shown together with the data at indicated $Q^2$ values. The full line shows the fit in the range $x<0.01$, the dotted line in the range $x<0.001$. The dots show the data with errors bars given by the statistical and systematical errors added in quadrature.  In the fit, all correlations of  errors were taken into account.}
\label{curv4} \end{figure}

\begin{figure}[htbp]
\centerline{
\includegraphics[width=9cm,angle=0]{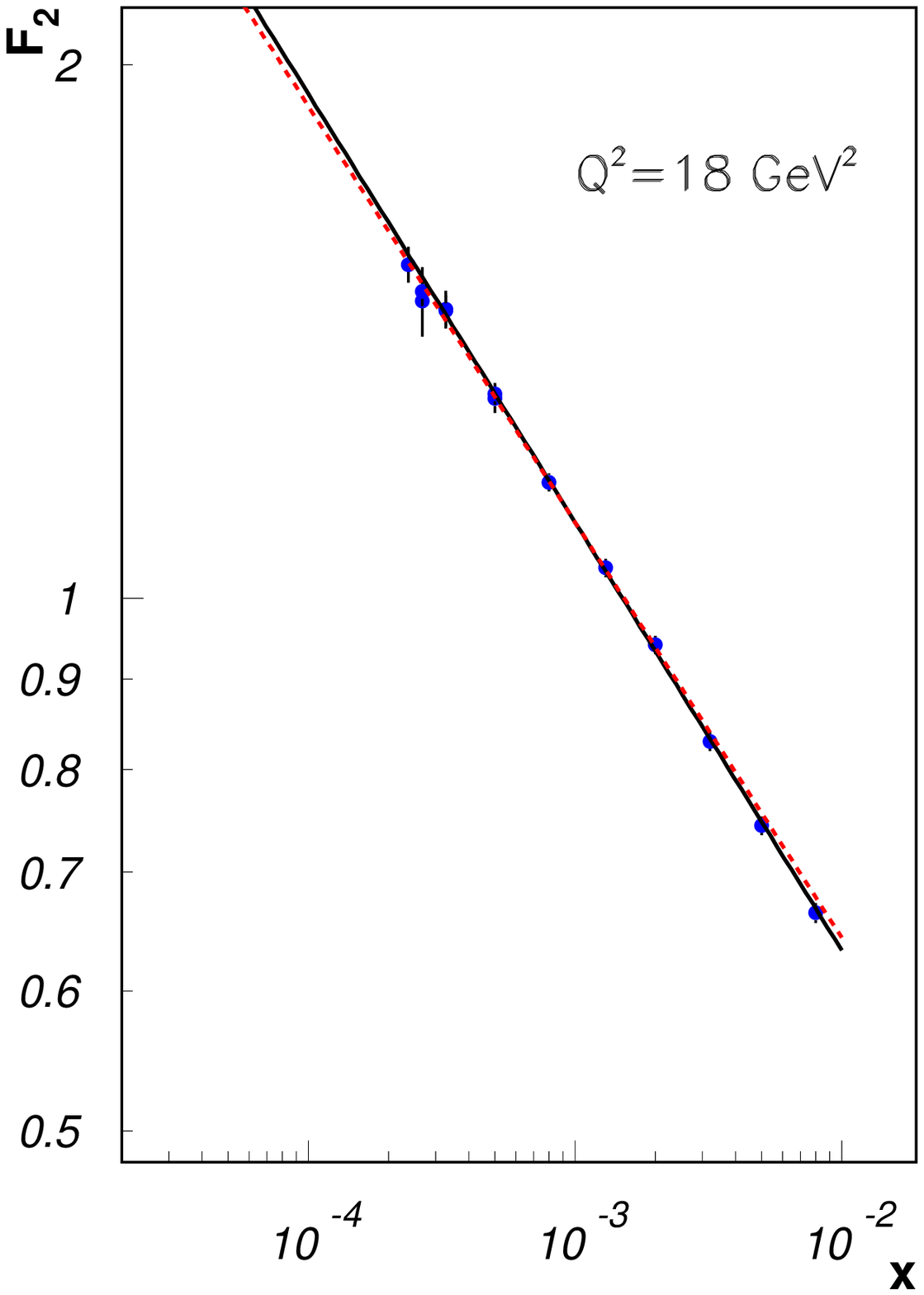}
\includegraphics[width=9cm,angle=0]{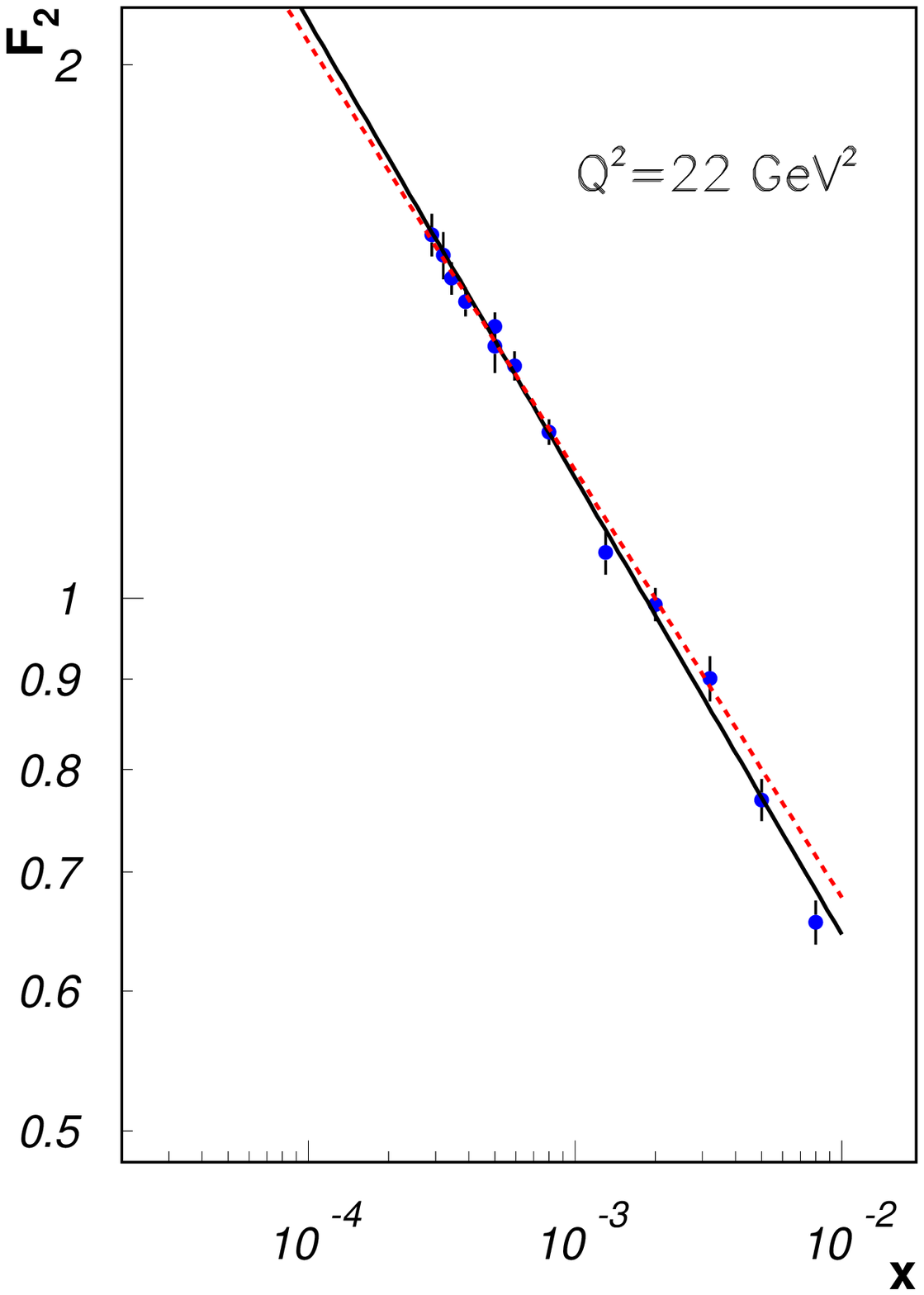}
}
\caption {The fitted curves, $F_2 \propto(1/x)^\lambda$, are shown together with the data at indicated $Q^2$ values. The full line shows the fit in the range $x<0.01$, the dotted line in the range $x<0.001$. The dots show the data with errors bars given by the statistical and systematical errors added in quadrature.  In the fit, all correlations of  errors were taken into account.}
\label{curv5} \end{figure}

\begin{figure}[htbp]
\centerline{
\includegraphics[width=9cm,angle=0]{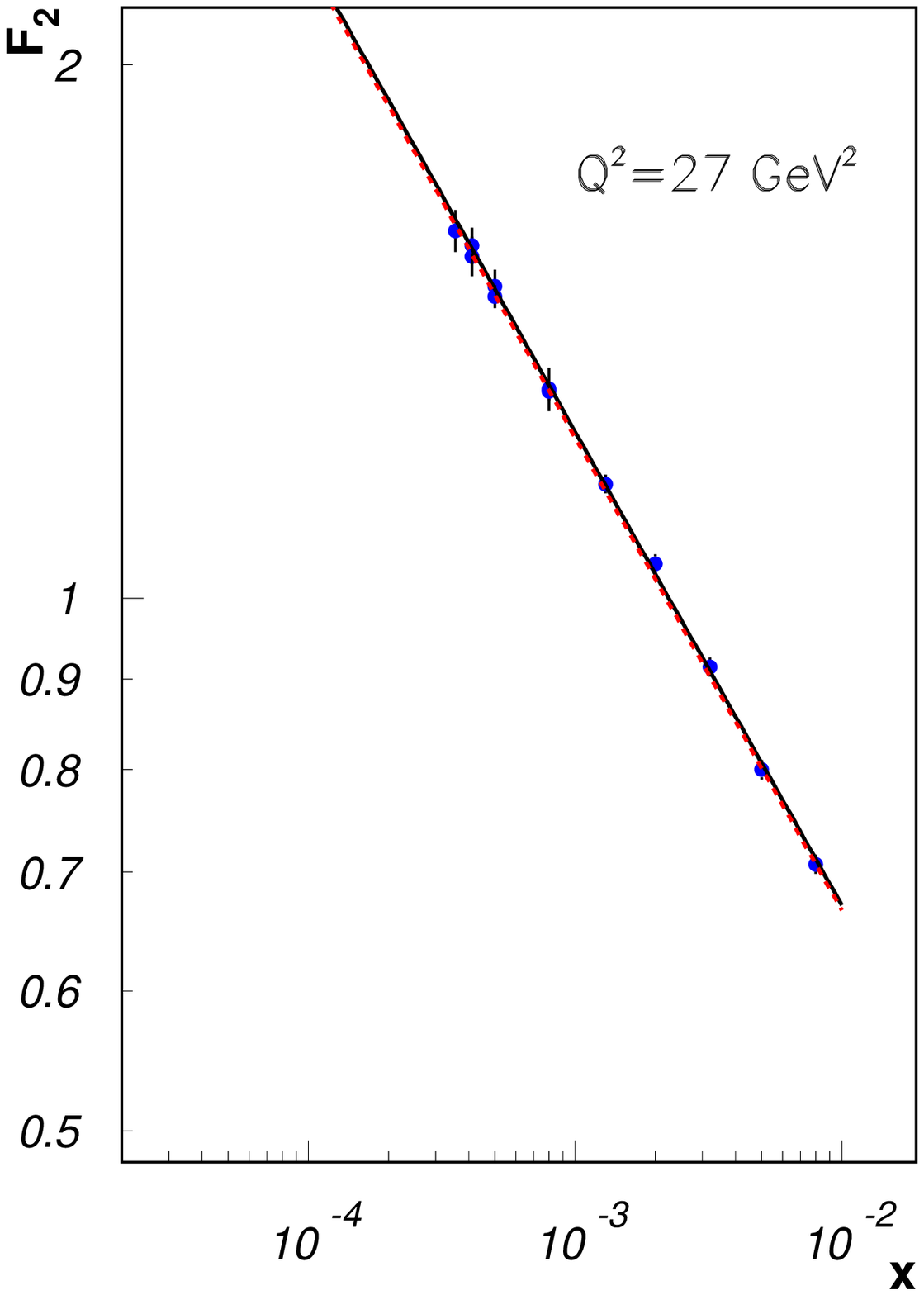}
\includegraphics[width=9cm,angle=0]{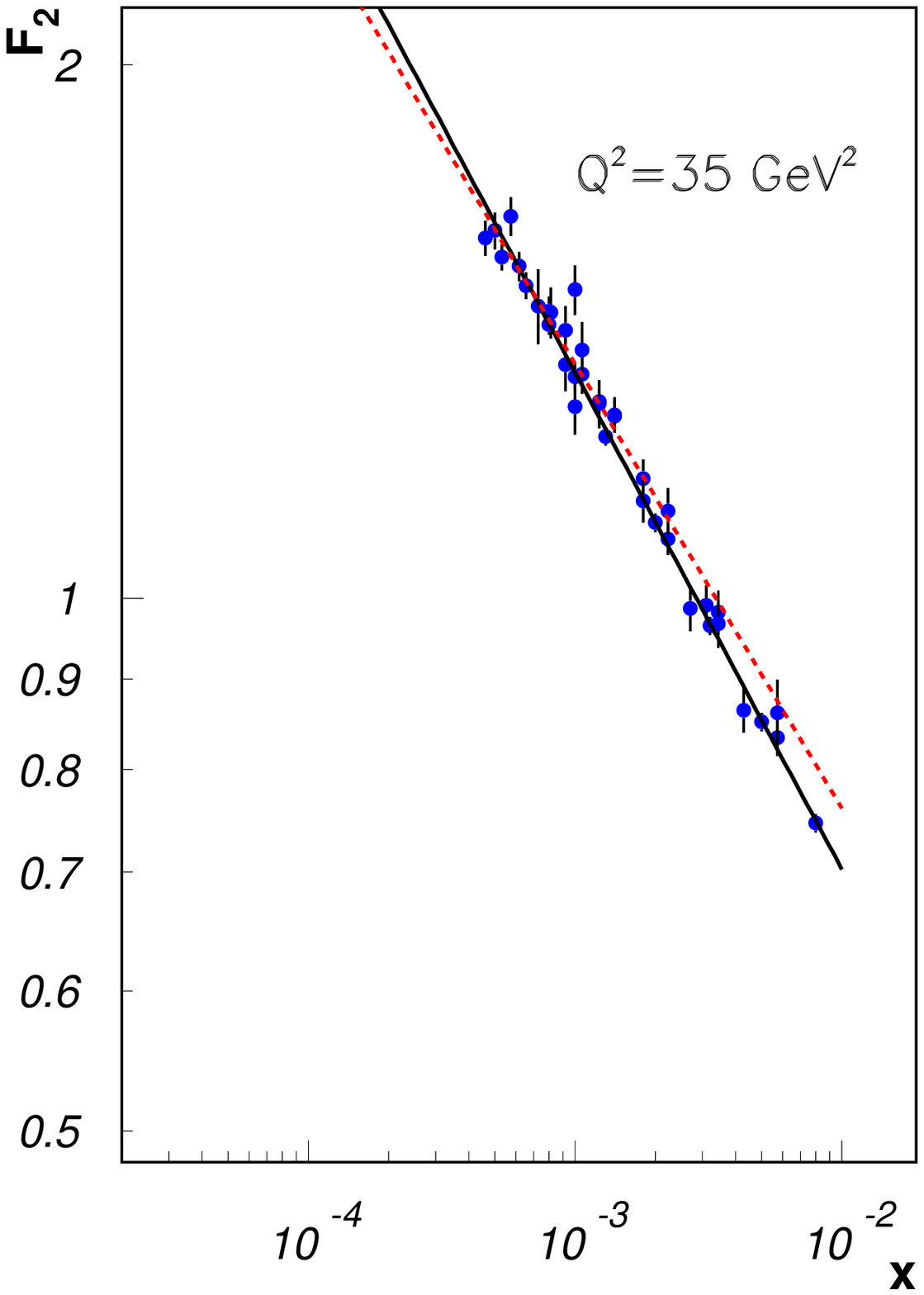}
}
\caption {The fitted curves, $F_2 \propto(1/x)^\lambda$, are shown together with the data at indicated $Q^2$ values. The full line shows the fit in the range $x<0.01$, the dotted line in the range $x<0.001$. The dots show the data with errors bars given by the statistical and systematical errors added in quadrature.  In the fit, all correlations of  errors were taken into account.}
\label{curv6} \end{figure}

\begin{figure}[htbp]
\centerline{
\includegraphics[width=9cm,angle=0]{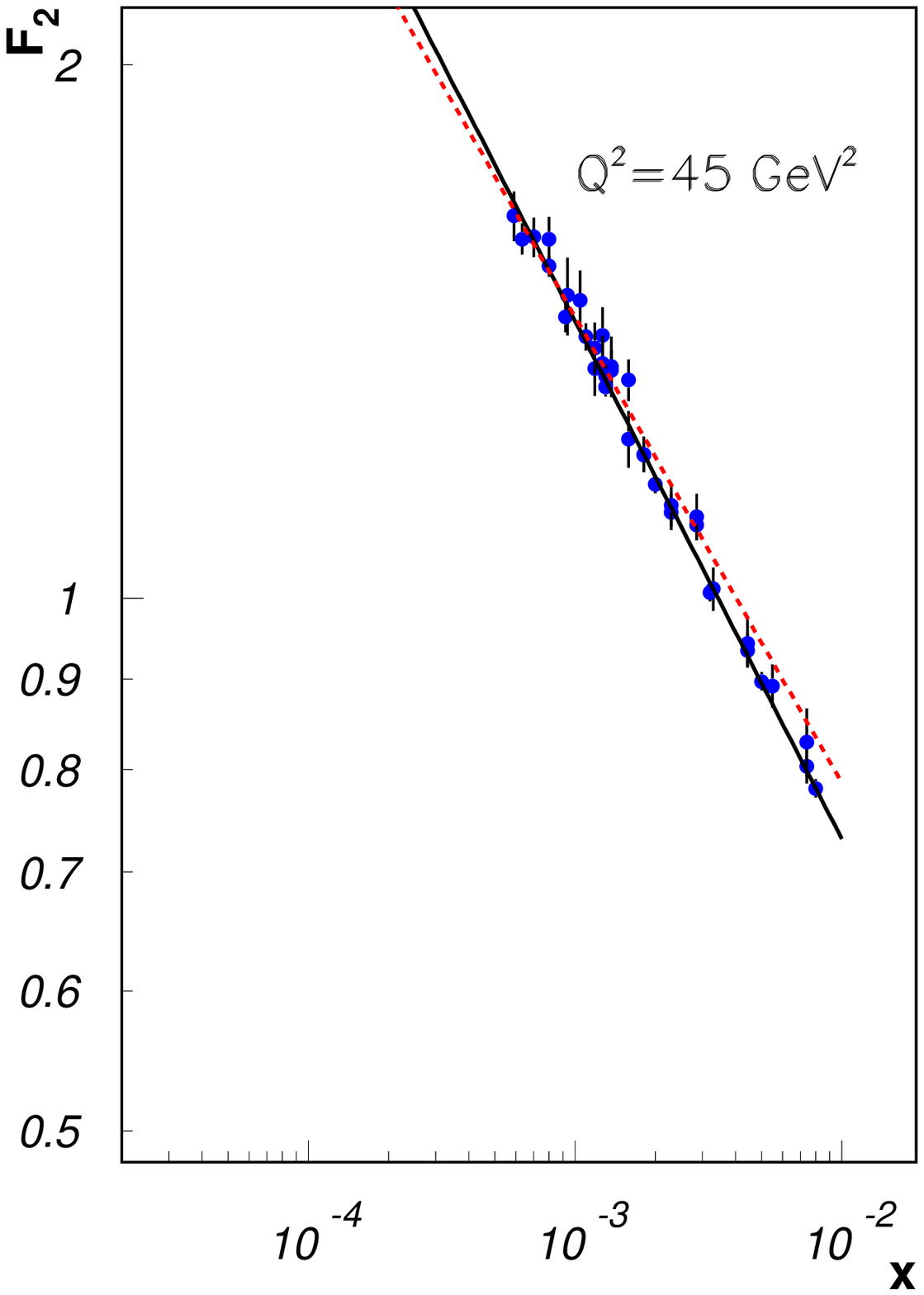}
}
\caption {The fitted curves, $F_2 \propto(1/x)^\lambda$, are shown together with the data at indicated $Q^2$ values. The full line shows the fit in the range $x<0.01$, the dotted line in the range $x<0.001$. The dots show the data with errors bars given by the statistical and systematical errors added in quadrature.  In the fit, all correlations of  errors were taken into account.}
\label{curv7} \end{figure}

In Fig.~\ref{curv1}, \ref{curv2}, \ref{curv3}, \ref{curv4}, \ref{curv5}, \ref{curv6} and  \ref{curv7}    we compare the fitted curves  to the corresponding data in all $Q^2$ regions, between  $2.7 $ and $ 45$ GeV$^2$.  The curves are drawn with the values of the constants  $\lambda$ and $A$  given in Tables 1 and 2. The full line shows the fit in the region $x<0.01$ and the dashed one the fit to the region $x<0.001$. The  curves are drawn in a larger region than the fit are performed, to emphasise the systematic differences between them.  The figures show quite clearly that the dashed lines are almost always below the full lines in the region of very small $x<0.001$. On the opposite end, around $x \approx 0.01$, the full line is almost always above the dashed one. Since the figures are  drawn on a double logarithmic scale (and are relatively large) it is possible to see, even clearly in some cases, a slight "bowing" of data due to a systematic decrease of  $\lambda$
at smaller $x$ values, seen in Fig.~\ref{lamcomp} and Table 1.

\begin{figure}[htbp]
\centerline{
\includegraphics[width=13cm,angle=0]{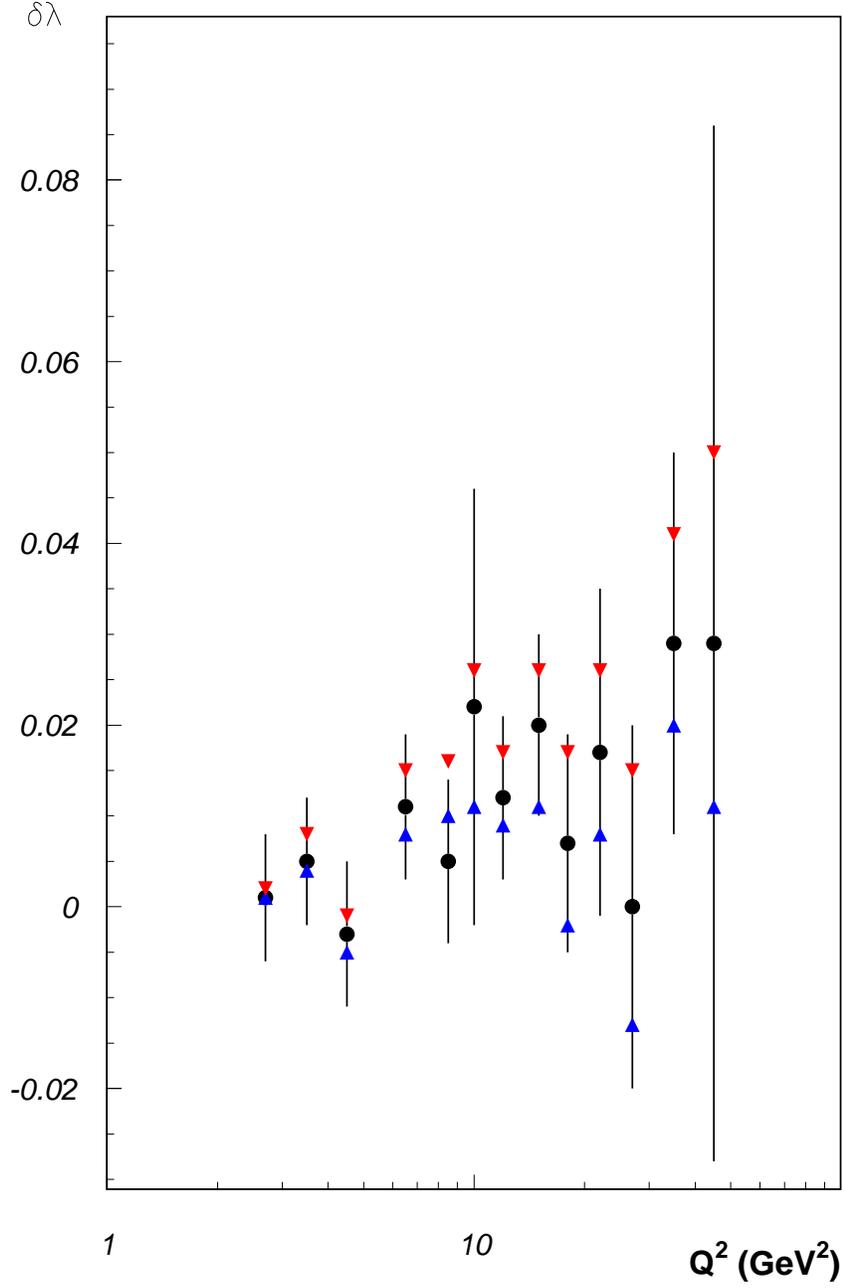}
}
\caption { Plot of  differences between the values of $\lambda $ determined  in the low-$x$ and very-low-$x$  region, $\Delta_\lambda = \lambda_{x <0.01}  -\lambda_{x<0.001} $.  The black dots show the results for $R=0.25$, the triangles pointing down show the results for $R=0.2$ and  the triangles pointing up show the results for $R=0.3$. The errors of black dots were obtained from  the quadratic addition of errors,  
$\delta_{\Delta_{\lambda}} =  \sqrt{\delta_{\lambda_{x <0.01}}^2  +\delta_{ \lambda_{x<0.001}}^2} $. The errors  of the triangles are of about the same magnitude and were not displayed, for clarity.  }
\label{lamcompsys} \end{figure}

To check the dependence of our results on the assumed value of $R$, we show in Fig.~\ref{lamcompsys} the  differences between the values of $\lambda $ determined  in the low $x$ and very low $x$  region, $\Delta_\lambda = \lambda_{x <0.01}  -\lambda_{x<0.001} $,
 assuming   $R=0.2$, $R=0.25$  and  $R=0.3$.  Fig.~\ref{lamcompsys}  shows that the rate of rise $\lambda$ remains sensitive to the assumed value of R, in spite of the cut on the variable $y=0.8$. Since the  value of  $R$ was determined as $R=0.23 \pm 0.04$  \cite{H1R2}, the main result of this paper remains valid
  even in the extreme  case of $R=0.3$.

\section{Discussion and Conclusions}

We have confirmed that the simple phenomenological function, $F_2 \propto(1/x)^\lambda$, fits very well the  high precision HERA data in the low $x$ region, $x<0.01$, at all $Q^2$ values between $ 0.3 < Q^{2} < 250 $ GeV$^2$.  Moreover, we have shown that this is also true for the very low $x$ region, $x<0.001$.
The new result of this investigation is that  the rate of rise, $\lambda$, determined in the two regions, indicate that $\lambda$ is systematically smaller in the very low $x$ region as compared to the low $x$ region, for $Q^2>6$ GeV$^2$. This result is obtained due to the high precision of the latest HERA data  \cite{hiH1ZEUS} whereas the earlier investigations concluded that the power $\lambda$  is  not dependent on $x$, within the experimental errors, see e.g.  ref.  \cite{h1lam}.   In a more recent investigation,   \cite{H1R}, it was  noticed, although indirectly,  that $\lambda$ values may diminish with decreasing $x$, in agreement with the present  investigation.

This observation may indicate the onset of BFKL behaviour in the very low $x$ region, $x<0.001$, $Q^2>6$ GeV$^2$, in agreement with the analysis of ref. \cite{KLRS17}. 
The analysis has shown that the BFKL gluon density describe data very well in this region and that the dominant contribution is provided by the second pole, which has $\omega \approx 0.14$, a value which is substantially smaller than the $\lambda $ value observed in this region, see Table 1. 
Note that the evaluation of ref. \cite{KLRS17} does not allow a quantitive  estimate of the expected change of the power $\lambda$ between the low $x$ and very low $x$ regions, as  the high quality of BFKL description  was obtained only in the second region.  

The observation of a systematic decrease of $\lambda$ with diminishing $x$  could also indicate that the double asymptotic region, in which $F_2 \approx \exp c \sqrt{\ln (1/x)}$, may be only few decades away from the $x<0.001$ region observed at HERA.   

In any case, our observation that the value of the exponent $\lambda$ decreases at small values of $x$, indicates that measurements at the future ep colliders, like VHEeP or LHeC \cite{Cald-Wing,LHeC}  will become  exciting, as they will approach the high energy limit of the virtual photon-hadron cross sections, where DGLAP and BFKL meets \cite{KLR16} and the confinement effects should become simple \cite{Feynman}.   

\section*{Acknowlegments}
We are grateful to Allen Caldwell and Sasha Glazov for very useful discussions and suggestions for improvements.

\begin{table}[ht]
\begin{center}
\begin{tabular}{|c|c|c|c|c|c|c|c|c|c|c|c|} 
\hline 
\hline 
& $x<0.01$ & & & &$x<0.001$ & & &\\
\hline 
$Q^2$& $\lambda$ & $\delta \lambda$ & $N_{df}$ & $\chi^2/N_{df}$& $\lambda$ & $\delta \lambda$& $N_{df}$ & $\chi^2/N_{df}$\\
\hline
\hline 
$0.35$  & 0.110 & 0.008 & 10 & 0.850 & 0.110 & 0.0081 & 10 & 0.850 \\
\hline
$0.4$ & 0.082 & 0.009 & 7 & 0.915 & 0.082 & 0.009 & 7 & 0.915\\
\hline
$0.5$ & 0.100 & 0.009 & 7 & 0.768 & 0.100& 0.009 & 7 & 0.768\\
\hline
$0.65$ & 0.121 & 0.011 & 7 & 0.813 & 0.121 & 0.011& 7 & 0.813\\
\hline
$0.85$ & 0.150 & 0.014 & 7 & 0.759 & 0.150 & 0.014& 7 & 0.759 \\
\hline
$1.2$ & 0.133 & 0.013 & 8 & 2.074& 0.133 & 0.013& 8 & 2.074 \\
\hline
$1.5$ & 0.142 & 0.009 & 10 & 1.741 & 0.142 & 0.009 & 10 & 1.741\\ 
\hline
$2$ & 0.159 & 0.007 & 10 & 1.246 & 0.159 & 0.007& 10 & 1.246 \\
\hline
$2.7$ & 0.169 & 0.005& 12 & 1.745 & 0.168 & 0.007 & 10 & 1.462 \\
\hline
$3.5$  & 0.173 & 0.004 & 18 & 1.391 & 0.168 & 0.007 & 16 & 1.485\\ 
\hline
$4.5$ & 0.189 & 0.004 & 13 & 1.908 & 0.192 & 0.008 & 10 & 2.347 \\
\hline
$6.5$ & 0.200 & 0.003 & 29 & 0.990 & 0.189 & 0.008 & 25 & 1.002\\ 
\hline
$8.5$  & 0.208 & 0.004 & 34 & 1.252 & 0.192 & 0.009 & 27 & 1.047 \\
\hline
$10$  & 0.226 & 0.007& 5 & 1.005 &0.205 & 0.024& 2& 0.518 \\
\hline
$12$ & 0.215 & 0.005 & 33 & 1.016 & 0.202 & 0.009& 24 & 0.715 \\
\hline
$15$  & 0.237 & 0.003 & 32 & 1.217 & 0.219 & 0.010  & 21 & 1.303 \\
\hline
$18$ & 0.242 & 0.003 & 11 & 0.401 & 0.234 & 0.012  & 6 & 0.358\\
\hline
$22$ & 0.258 & 0.007  & 11 & 0.961 & 0.241 & 0.018& 6 & 0.601 \\
\hline
$27$ & 0.267 & 0.004 & 10 & 0.632 & 0.267 & 0.020 & 5 & 0.136 \\
\hline
$35$ & 0.280 & 0.003 & 35 & 1.144 & 0.251 & 0.021 & 13 & 1.759 \\
\hline
$45$ & 0.292 & 0.004 & 33 & 0.877 & 0.263& 0.057 & 5 & 1.085\\
\hline
$60$ & 0.313 & 0.005 & 32 & 1.274 & & &  & \\
\hline
$70$ &  0.332 & 0.009 & 10 & 0.812& & &  & \\
\hline
$90$ & 0.321 & 0.007 & 27 & 0.925 & & &  & \\
\hline
$120$ & 0.352 & 0.008& 31 & 0.506& & &  &\\
\hline
$150$ & 0.339 & 0.011 & 17 & 1.101 & & &  &\\
\hline
$200$ & 0.373 & 0.010 & 16 & 0.545& & &  &\\
\hline
$250$ & 0.417 & 0.014 & 13 & 0.727& & &  &\\
\hline

\end{tabular}
\caption{ The results for the $\lambda$ constant obtained from the fits of the function $F_2 =  A\, (1/x)^\lambda$  to the final, high precision H1 and ZEUS data at all $Q^2$ values, in the low-$x$ and very-low-$x$ regions, $x<0.01$ and  $x<0.001$. Note that below $Q^2<2.7 $ GeV$^2$ there are no differences between the two regions.
 }
\end{center}
\end{table}

\newpage

\begin{table}[ht]
\begin{center}
\begin{tabular}{|c|c|c|c|c|c|c|c|c|c|c|c|} 
\hline 
\hline 
& $x<0.01$ & & & &$x<0.001$ & & &\\
\hline 
$Q^2$& $A$ & $\delta A$ & $N_{df}$ & $\chi^2/N_{df}$& $A$ & $\delta A$& $N_{df}$ & $\chi^2/N_{df}$\\
\hline
\hline 
$0.35$  & 0.0957 & 0.008 & 10 & 0.850 & 0.0957 & 0.008 & 10 & 0.850 \\
\hline
$0.4$ & 0.143 & 0.013 & 7 & 0.915 & 0.143 & 0.014 & 7 & 0.915\\
\hline
$0.5$ & 0.136 & 0.013 & 7 & 0.768 & 0.136& 0.013 & 7 & 0.768\\
\hline
$0.65$ & 0.133 & 0.015 & 7 & 0.813 & 0.133 & 0.015 & 7 & 0.813\\
\hline
$0.85$ & 0.120 & 0.018 & 7 & 0.759 & 0.120 & 0.018 & 7 & 0.759 \\
\hline
$1.2$ & 0.170 & 0.022 & 8 & 2.074& 0.170 & 0.022 & 8 & 2.074 \\
\hline
$1.5$ & 0.180 & 0.016 & 10 & 1.741 & 0.184 & 0.016 & 10 & 1.741\\ 
\hline
$2$ & 0.172 & 0.012 & 10 & 1.246 & 0.172 & 0.012 & 10 & 1.246 \\
\hline
$2.7$ & 0.179 & 0.008& 12 & 1.745 & 0.181 & 0.012 & 10 & 1.462 \\
\hline
$3.5$  & 0.195 & 0.007 & 18 & 1.391 & 0.203 & 0.012 & 16 & 1.485\\ 
\hline
$4.5$ & 0.191 & 0.007 & 13 & 1.908 & 0.186 & 0.0013 & 10 & 2.347 \\
\hline
$6.5$ & 0.201 & 0.006 & 29 & 0.990 & 0.221 & 0.015 & 25 & 1.002\\ 
\hline
$8.5$  & 0.211 & 0.007 & 34 & 1.252 & 0.234 & 0.016 & 27 & 1.047 \\
\hline
$10$  & 0.192 & 0.010  & 5 & 1.005 &  0.226    &  0.044  & 2& 0.518 \\
\hline
$12$ & 0.225 & 0.008 & 33 & 1.016 & 0.247 & 0.018& 24 & 0.715 \\
\hline
$15$  & 0.202 & 0.005 & 32 & 1.217 & 0.232 & 0.019  & 21 & 1.303 \\
\hline
$18$ & 0.208 & 0.005 & 11 & 0.401 & 0.218 & 0.021  & 6 & 0.358\\
\hline
$22$ & 0.197 & 0.010  & 11 & 0.961 & 0.223 & 0.030 & 6 & 0.601 \\
\hline
$27$ & 0.196 & 0.005 & 10 & 0.632 & 0.195 & 0.029 & 5 & 0.136 \\
\hline
$35$ & 0.193 & 0.005 & 35 & 1.144 & 0.239 & 0.037 & 13 & 1.759 \\
\hline
$45$ & 0.190 & 0.005 & 33 & 0.877 & 0.234& 0.096 & 5 & 1.085\\
\hline
$60$ & 0.179 & 0.005 & 32 & 1.274 & & &  & \\
\hline
$70$ &  0.164& 0.009 & 10 & 0.812& & &  & \\
\hline
$90$ & 0.188 & 0.008 & 27 & 0.925 & & &  & \\
\hline
$120$ & 0.163 & 0.008  & 31 & 0.506& & &  &\\
\hline
$150$ & 0.183 & 0.011 & 17 & 1.101 & & &  &\\
\hline
$200$ & 0.158 & 0.008 & 16 & 0.545& & &  &\\
\hline
$250$ & 0.129 & 0.010 & 13 & 0.727& & &  &\\
\hline

\end{tabular}
\caption{  The results for the $A$ constant obtained from the fits of the function $F_2 = A\, (1/x)^\lambda$  to the final, high precision H1 and ZEUS data at all $Q^2$ values, in the low-$x$ and very-low-$x$ regions, $x<0.01$ and  $x<0.001$. Note that below $Q^2<2.7$ GeV$^2$ there are no differences between the two regions.
 }
\end{center}
\end{table}

\newpage

\end{document}